\documentstyle[amssymb,aps,pra,12pt]{revtex}

\begin{document}
\title{Scheme for demonstrating Bell theorem in tripartite entanglement between
atomic ensembles}
\author{Xi-Bin Zhou, Yong-Sheng Zhang\thanks{%
Electronic address: yshzhang@ustc.edu.cn} and Guang-Can Guo\thanks{%
Electronic address: gcguo@ustc.edu.cn}}
\address{Laboratory of Quantum Information, University of Science and Technology of\\
China, CAS, Hefei 230026, People's Republic of China\bigskip \bigskip }
\maketitle

\begin{abstract}
\baselineskip24pt We propose an experimentally feasible scheme to
demonstrate quantum nonlocality, using Greenberger-Horne-Zeilinger (GHZ) and 
$W$ entanglement between atomic ensembles generated by a new developed
method based on laser manipulation and{} single-photon detection.

PACS number(s): 03.65.Ud, 03.67.-a, 42.50.Gy\newpage
\end{abstract}

\baselineskip24pt $~\;\quad \qquad $

\section{Introduction}

Quantum entanglement, which is so contradictory to our intuition, maybe the
most fundamental feature of quantum mechanics. From the paper of Einstein,
Podolsky, and Rosen (EPR) in 1935\cite{EPR}, quantum entanglement has been
carefully investigated. It has found wide applications in the demonstration
of quantum nonlocality and quantum information processing, such as quantum
teleportation\cite{tele} and quantum cryptography\cite{cry}. Greenberger,
Horne, and Zeilinger have shown that quantum mechanical predictions for
certain measurement results on three entangled particles are in conflict
with local realism in cases where quantum theory makes definite predictions,
whereas as for EPR state with two entangled particles, the conflict with
local realism only results from statistical predictions\cite{ghz}. Actually,
there are only two classes of inequivalent tripartite entanglements under
local operation assisted by classical communication (LOCC)\cite{Dur}: one is
GHZ state, expressed as 
\begin{equation}
\frac 1{\sqrt{2}}(|000\rangle +|111\rangle );  \eqnum{1}
\end{equation}
the other is $W$ state, expressed as 
\begin{equation}
\frac 1{\sqrt{3}}(|001\rangle +|010\rangle +|100\rangle ).  \eqnum{2}
\end{equation}
The experiments to observe photons' GHZ state and demonstrate quantum
nonlocality by it have been accomplished\cite{6,7}. In these experiments,
the subsystem of the entanglement is single photon.

Recently, a new method to generate entanglement between atomic ensembles has
been developed\cite{Duan,Duan0208,Duan2,Xue}. In this paper we adopt this
method to generate GHZ and $W$ state and devise experimental schemes to
demonstrate quantum nonlocality using the generated entanglement states. In
contrast to the existing schemes, the one described here takes advantage of
the virtues of entanglement between atomic ensembles, such as long lived
time and more robust resilience to realistic imperfection..

\section{GHZ state}

The basic element of our scheme is an ensemble of many identical alkali
atoms, which can be experimentally realized as either a room-temperature
atomic gas\cite{9,10} or a sample of cold trapped atoms \cite{11,12}. The
relevant level structure of the atom is shown in Fig. 1.

\begin{center}
\smallskip {\bf Figure 1}
\end{center}

A pair of metastable states $|g\rangle $ and $|s\rangle $ can correspond
to---for example---hyperfine or Zeeman sublevels of the atom. From the two
levels $|g\rangle $ and $|s\rangle $ we can define a collective atomic
operator $S=(1/\sqrt{N_a})\sum_i|g\rangle _i\langle s|$ where $N_a\gg 1$ is
the total atom number. The ensemble ground state can be expressed as $%
|0_a\rangle =\prod_i|g\rangle _i$. Here we use the same symbols as in Ref.%
\cite{Duan}.

The process of the preparation scheme in Ref.\cite{Duan} is simply described
as follows: The ensembles L and R are initially prepared in the ground state
and then excited respectively by a short Raman pulse applied to the
transition $|g\rangle \rightarrow |e$ $\rangle $. The Raman pulse is so weak
that the forward--scattered Stokes light from the transition $|e\rangle
\rightarrow |s\rangle $ has a mean number much smaller than 1. The
forward--scattered Stokes lights from the two ensembles are then interfered
at a beam splitter and further detected by two single-photon detectors. In
the cases where only one detector register a click, we can not distinguish
from which ensemble this registered photon comes; due to this
indistinguishability, the projected state of the ensembles L and R is nearly
maximally entangled, with the form 
\begin{equation}
|\Psi _{LR}\smallskip \rangle =\frac 1{\sqrt{2}}(S_L^{\dagger }+e^{i\phi
}S_R^{\dagger })|vac\rangle ,  \eqnum{3}
\end{equation}
where $\phi $ is an unknown phase difference fixed by the optical channel
connecting the L and R ensembles, and $|vac\rangle =|0_a\rangle
_L|0_a\rangle _R.$

We can generate GHZ state by three pairs of such atomic ensembles. The whole
state can be described by 
\begin{equation}
|\Psi \rangle =\prod_{i=1}^3[\frac 1{\sqrt{2}}(S_{L_i}^{\dagger }+e^{i\phi
_i}S_{R_i}^{\dagger })]|vac\rangle ,  \eqnum{4}
\end{equation}
where $|vac\rangle $ denote the vacuum of the whole six ensembles. In the
expansion of the state (4), there are only two components which have one
excitation in each pair. This component state is given by\qquad \qquad
\qquad \qquad \qquad 
\begin{equation}
|\Psi _{GHZ}\rangle =(1/\sqrt{2})(\prod_{i=1}^3S_{L_i}^{\dagger }+e^{i\phi
_r}\prod_{i=1}^3S_{R_i}^{\dagger })|vac\rangle ,  \eqnum{5}
\end{equation}
with $\phi _r=\phi _1+\phi _2+\phi _3$, which is exactly the three-party GHZ
maximal entangled states in the `polarization' basis. In the system, by
applying retrieval pulses of suitable polarization that are near-resonant
with the atomic transition $|s\rangle \rightarrow |e\rangle $, we can
simultaneously convert the stored atomic excitations into light, and by
using single-bit rotations, such as Hadamard transformations, and the number
detection through single-photon detectors, we can generate the GHZ state
which is practical for experiment. The method described above is similar to
the scheme proposed in Ref.\cite{Duan2} to entangle many atomic ensembles,
but there are some differences in detail.

The efficiency of our scheme can be described by the total time needed to
register the effective GHZ state. Through the similar analysis to Duan's\cite
{Duan2}, we can know the total time for registering the GHZ state is $%
T\thicksim 4t_0/(1-\eta )^3$, where $t_0$ is the preparation time of state
(3), and $\eta $ describe the overall loss probability of photon detectors.

Now we can use the entanglement state generated above to test the quantum
nonlocality. The scheme we present here is similar to the scheme in Ref.\cite
{7} in principle, but because we use a different entanglement source, which
is a `polarization' maximal entangled (PME) \cite{Duan} state between atomic
ensembles, we should adjust the measurement manners. The diagram of our
set-up is shown in Fig. 2.

\qquad \qquad \qquad \qquad \qquad \qquad \qquad \qquad {\bf Figure 2}

Analogous to polarization entanglement, we can write the generated photons'
GHZ state as

\begin{equation}
|\Psi _{GHZ}\rangle =\frac 1{\sqrt{2}}(|L_1\rangle |L_2\rangle |L_3\rangle
+|R_1\rangle |R_2\rangle |R_3\rangle ),  \eqnum{6}
\end{equation}
where $|L_i$ $\rangle $ denotes that a photon is emitted from the atomic
ensemble $L_i$, and $|R_i\rangle $ denotes that a photon is emitted from the
atomic ensemble $R_{i+1}$, here we assume the notation 3+1$\equiv $1 for the
subscripts. It is obvious that the three photons are in a quantum
superposition of the state $|L_1\rangle |L_2\rangle |L_3\rangle $ (all three
photons are emitted from the atomic ensembles marked by $L$)$,$ and the
state $|R_1\rangle |R_2\rangle |R_3\rangle $ (all three photons are emitted
from the atomic ensembles marked by $R$), so none of the three photons has a
well defined state of its own.

We can also consider measurements of other bases which can be expressed as 
\begin{eqnarray}
|L^{\prime }\rangle &=&\frac 1{\sqrt{2}}(|L\rangle +|R\rangle ),  \eqnum{7}
\\
|R^{\prime }\rangle &=&\frac 1{\sqrt{2}}(|L\rangle -|R\rangle );  \nonumber
\\
|L^{\prime \prime }\rangle &=&\frac 1{\sqrt{2}}(|L\rangle +i|R\rangle ), 
\nonumber \\
|R^{\prime \prime }\rangle &=&\frac 1{\sqrt{2}}(|L\rangle -i|R\rangle ). 
\nonumber
\end{eqnarray}
For convenience we will refer to a measurement of $L^{\prime }/R^{\prime }$
as a $\sigma _x$ measurement, and one of $L^{\prime \prime }/R^{\prime
\prime }$ as a $\sigma _y$ measurement.

It has been shown that the demonstration of the conflict between quantum
mechanics and local realism consists of four experiments, each with three
spatially separated `polarization' measurements\cite{7}. First, we perform $%
\sigma _y\sigma _y\sigma _x,$ $\sigma _y\sigma _x\sigma _y$ and $\sigma
_x\sigma _y\sigma _y$ experiments. If we obtained the results predicted for
the GHZ state\smallskip , then for a $\sigma _x\sigma _x\sigma _x$
experiment, our consequent expectations according to local-realism are
exactly the opposite of the expectations according to quantum mechanics\cite
{ghz}. In our set-up shown by Fig. 2, when none phase plate is installed we
implement a $\sigma _x$ measurement, and when a $\frac \pi 2$ phase plate is
installed, we implement a $\sigma _y$ measurement. By observing which
detector register a photon, we can know the result of a measurement. So we
can test the quantum nonlocality in the same steps as in Ref.\cite{7}.\qquad
\qquad \qquad \qquad \qquad \qquad \qquad \qquad \qquad \qquad \qquad \qquad
\qquad \qquad \qquad \qquad \qquad \qquad \qquad \qquad \qquad \qquad \qquad
\qquad \qquad \qquad \qquad \qquad \qquad \qquad \qquad \qquad \qquad \qquad
\qquad \qquad \qquad \qquad \qquad \qquad \qquad \qquad \qquad \qquad \qquad
\qquad \qquad \qquad \qquad \qquad \qquad \qquad \qquad \qquad \qquad \qquad
\qquad \qquad \qquad \qquad \qquad \qquad \qquad \qquad \qquad \qquad \qquad
\qquad \qquad \qquad \qquad \qquad \qquad \qquad \qquad \qquad \qquad \qquad
\qquad \qquad \qquad \qquad \qquad \qquad \qquad \qquad \qquad \qquad \qquad
\qquad \qquad \qquad \qquad \qquad \qquad \qquad \qquad \qquad \qquad \qquad
\qquad \qquad \qquad \qquad \qquad \qquad \qquad \qquad \qquad \qquad \qquad
\qquad \qquad \qquad \qquad \qquad \qquad \qquad \qquad \qquad \qquad \qquad
\qquad \qquad \qquad \qquad \qquad \qquad \qquad \qquad \qquad \qquad \qquad
\qquad \qquad \qquad \qquad \qquad \qquad \qquad \qquad \qquad \qquad \qquad
\qquad \qquad \qquad \qquad \qquad \qquad \qquad \qquad \qquad \qquad \qquad
\qquad \qquad \qquad \qquad \qquad \qquad \qquad \qquad \qquad \qquad \qquad
\qquad \qquad \qquad \qquad \qquad \qquad \qquad \qquad \qquad \qquad \qquad
\qquad \qquad \qquad \qquad \qquad \qquad \qquad \qquad \qquad \qquad \qquad
\qquad \qquad \qquad \qquad \qquad \qquad \qquad \qquad \qquad \qquad \qquad
\qquad \qquad \qquad \qquad \qquad \qquad \qquad \qquad \qquad \qquad \qquad
\qquad \qquad \qquad \qquad \qquad \qquad \qquad \qquad \qquad \qquad \qquad
\qquad \qquad \qquad \qquad \qquad \qquad \qquad \qquad \qquad \qquad \qquad
\qquad \qquad \qquad \qquad \qquad \qquad \qquad \qquad \qquad \qquad \qquad
\qquad \qquad \qquad \qquad \qquad \qquad \qquad \qquad \qquad \qquad \qquad
\qquad \qquad \qquad \qquad \qquad \qquad \qquad \qquad \qquad \qquad \qquad
\qquad \qquad \qquad \qquad \qquad \qquad \qquad \qquad \qquad \qquad \qquad
\qquad \qquad \qquad \qquad \qquad \qquad \qquad \qquad \qquad \qquad \qquad
\qquad \qquad \qquad \qquad \qquad \qquad \qquad \qquad \qquad \qquad \qquad
\qquad \qquad \qquad \qquad \qquad \qquad \qquad \qquad \qquad \qquad \qquad
\qquad \qquad \qquad \qquad \qquad \qquad \qquad \qquad \qquad \qquad \qquad
\qquad \qquad \qquad \qquad \qquad \qquad \qquad \qquad \qquad \qquad \qquad
\qquad \qquad \qquad \qquad \qquad \qquad \qquad \qquad \qquad \qquad \qquad
\qquad \qquad \qquad \qquad \qquad \qquad \qquad \qquad \qquad \qquad \qquad
\qquad \qquad \qquad \qquad \qquad \qquad \qquad \qquad \qquad \qquad \qquad
\qquad \qquad \qquad \qquad \qquad \qquad \qquad \qquad \qquad \qquad \qquad
\qquad \qquad \qquad \qquad \qquad \qquad \qquad \qquad \qquad \qquad \qquad
\qquad \qquad \qquad \qquad \qquad \qquad \qquad \qquad \qquad \qquad \qquad
\qquad \qquad \qquad \qquad \qquad \qquad \qquad \qquad \qquad \qquad \qquad
\qquad \qquad \qquad \qquad \qquad \qquad \qquad \qquad \qquad

\section{$W$ state}

\smallskip In the following we propose a similar method to generate $W$
state, which is also based on the preparation of the state (3). The method
is divided into three steps as follows:

\smallskip First, we prepare three pairs of atomic ensembles in the state
(3) 
\begin{equation}
|\Psi _i\rangle =[\frac 1{\sqrt{2}}(S_{B_i}^{\dagger }+e^{i\phi
_i}S_{C_i}^{\dagger })/]|vac\rangle \text{ }(i=1,2,3).  \eqnum{8}
\end{equation}

Second, we prepare non-PME $W$ state between three atomic ensembles denoted
by $A_1$, $A_2$ and $A_3$. The set-up is shown in Fig. 3.

\qquad \qquad \qquad \qquad \qquad \qquad \qquad \qquad {\bf Figure 3}

All the atoms of the three ensembles are initially prepared in the ground
state $|g\rangle $. The three ensembles are put in a line, and illuminated
by a short, off-resonant laser pulse that induces Raman transition into the
state $|s\rangle $. Behind the third ensemble, we put a filter which can
eliminate the pump laser pulse from the forward-scattered Stokes photon, and
a single-photon detector to detect the Stokes photon. The set-up is shown in
Fig. 3. Because in the free space the velocity of the Stokes light is close
to $c$, so the delay between the Stokes photons emitted from different
ensembles is much smaller than the pulse width. Therefore when we register
only one click in the detector, we can not distinguish which ensemble this
registered photon comes from, so we obtain the state\qquad \qquad \qquad
\qquad \qquad 
\begin{equation}
|\Psi _A\rangle =\frac 1{\sqrt{3}}(S_{A_1}^{\dagger }+e^{i\phi
_{A2}}S_{A_2}^{\dagger }+e^{i\phi _{A3}}S_{A_3}^{\dagger })|vac\rangle , 
\eqnum{9}
\end{equation}
where $|vac\rangle $ denote the vacuum state of the three ensembles. We
should notice that $|\Psi _A\rangle $ is entangled in Fock basis which is
experimentally hard to do certain single-bit operations. The whole state of
the nine ensembles can be described by

\begin{equation}
|\Psi ^{\prime }\rangle =|\Psi _1\rangle |\Psi _2\rangle |\Psi _3\rangle
|\Psi _A\rangle .  \eqnum{10}
\end{equation}

Third, in the expansion of the state (10), there are only three components
which have one excitation in each pair of $A_i$ and $B_i$. This component
state is given by 
\begin{equation}
|\Psi _W\rangle =\frac 1{\sqrt{3}}(e^{i\phi _1}S_{A_1}^{\dagger
}S_{B_2}^{\dagger }S_{B_3}^{\dagger }S_{C_1}^{\dagger }+e^{i(\phi _{A2}+\phi
_2)}S_{A_2}^{\dagger }S_{B_1}^{\dagger }S_{B_3}^{\dagger }S_{C_2}^{\dagger
}+e^{i(\phi _{A3}+\phi _3)}S_{A_3}^{\dagger }S_{B_1}^{\dagger
}S_{B_2}^{\dagger }S_{C_3}^{\dagger }).  \eqnum{11}
\end{equation}

Using the set-up in Fig. 4, we apply retrieval pulses to ensembles $A_i$ and 
$B_i$ $(i=1,2,3)$ and register only the coincidence of the three party, i.e.
there is one and only one click on each party, through the postselection
techniques an experimentally practical $W$ state can be obtained.

\qquad \qquad \qquad \qquad \qquad \qquad \qquad \qquad {\bf Figure 4}

\smallskip If the state $|\Psi _i\rangle $ ($i=1,2,3)$ and $|\Psi _A\rangle $
can be prepared independently at the same time, the total preparation time
of state $|\Psi _W\rangle $ is $4\max (t_0,$ $t_1)/(1-\eta )^3,$ where $t_0$
is the preparation time of $|\Psi _i\rangle ,$ and $t_1$ is the preparation
time of state $|\Psi _A\rangle .$ In fact, we can easily know that when
atomic ensembles are illuminated by pumping light, the excitation
probability of state $|\Psi _i\rangle $ and state $|\Psi _A\rangle $ are at
the same order, as well as $t_0$ and $t_1$.

Now let we see how to demonstrate quantum nonlocality by $W$ state. Here we
use the same notations as in Ref.\cite{Cabello}: $z_i$ and $x_i$ will be the
results ($-1$ or $1$) of $\sigma _z$ and $\sigma _x$ measurements on the
party $i$ ($i=1,2,3$). By applying retrieval pulses, if a Stokes photon was
detected by the detector $D_A^i$, we assume $x_i=1,$ and if it is detected
by the detector $D_B^i$ $(i=1,2,3)$, we assume $x_i=-1$. We can implement a $%
\sigma _z$ measurement on party $i$ by applying a retrieval pulse to the
ensemble $C_i$, and if a Stokes photon is detected, i.e. the ensemble has an
excitation, we assume $z_i=1,$ if not we assume $z_i=-1.$

According to the rules described above, we can easily check the following
three properties experimentally.

\begin{eqnarray}
P(z_i &=&-1,\text{ }z_j=-1)=1,  \eqnum{12} \\
P(x_j &=&x_k|\text{ }z_i=-1)=1,  \nonumber \\
P(x_i &=&x_k|\text{ }z_j=-1)=1,  \nonumber
\end{eqnarray}
where $P(z_i=-1,$ $z_j=-1)$ means the probability of two parties (although
we cannot tell which two) giving the result $-1$ when implementing $\sigma
_z $ measurements on all three parties, and $P(x_j=x_k|$ $z_i=-1)$ is the
conditional probability of two $\sigma _x$ measurements on parties $i$ and $%
j $ having the same results given that the result of a $\sigma _z$
measurement on party $k$ is $-1$ $(i\neq j\neq k).$

\smallskip Based on EPR's local realism, we can deduced from the properties
(12) that $P(x_i=x_j=x_k)=1$, but according to quantum mechanics $%
P(x_i=x_j=x_k)=\frac 34,$ where $P(x_i=x_j=x_k)$ means the probability of
three separated $\sigma _x$ measurements giving the same results. So we can
demonstrate quantum nonlocality with $W$ state. We can also consider
Mermin's inequality with tripartite entanglement\cite{Mermin} 
\begin{equation}
-2\leqslant \langle a_1a_2a_3\rangle -\langle a_1b_2b_3\rangle -\langle
b_1a_2b_3\rangle -\langle b_1b_2a_3\rangle \leqslant 2,  \eqnum{13}
\end{equation}
where $a_i$ and $b_i$ are observables of qubit $i$. By choosing $a_i=\sigma
_{z_i}$ and $b_i=\sigma _{x_i}$, we can observe the violation of the
inequality by experiments.

\section{\protect\smallskip Conclusion}

\smallskip In summary we have proposed an experimental scheme to generate
GHZ and $W$ state between macroscopic atomic ensembles, and demonstrate
quantum nonlocality by the generated entanglement state. With the current
technology, we can realize the scheme which benefits much from the important
property of built-in entanglement purification, furthermore, all we use in
the scheme are linear optical elements, which make it easy to manipulate\cite
{Duan}. \smallskip In addition, our scheme can be easily generalized to
entangle more than three atomic ensembles or only two ensembles.

\begin{center}
{\bf Acknowledgments}
\end{center}

This work was funded by the National Fundamental Research Program
(2001CB309300), the Innovation Funds from Chinese Academy of Sciences, and
also by the outstanding Ph. D thesis award and the CAS's talented scientist
award rewarded to Lu-Ming Duan.

{\bf Figure Caption}.

{\bf Figure 1.} The relevant level structure of the atoms in the ensemble,
with $|g\rangle ,$ the ground state, $|s\rangle ,$ the metastable state for
storing a qubit, and $|e\rangle ,$ the excited state.

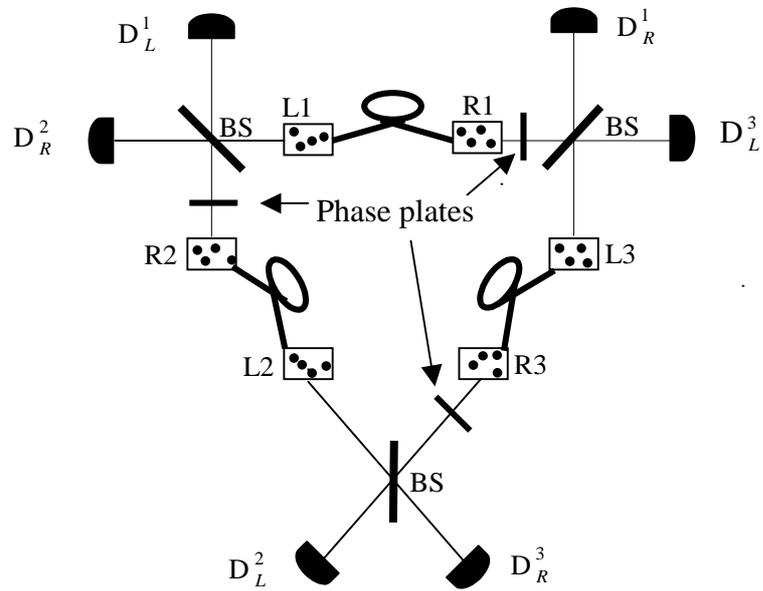
{\bf Figure 2.} Set-up for generating GHZ state between atomic ensembles and
demonstrating quantum nonlocality.

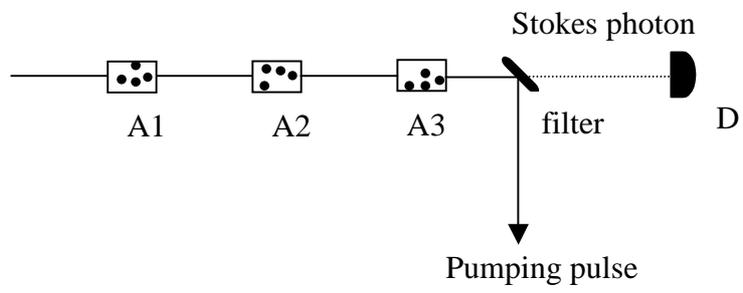
{\bf Figure 3. }Set-up for generating non-PME $W$ state between atomic
ensembles.

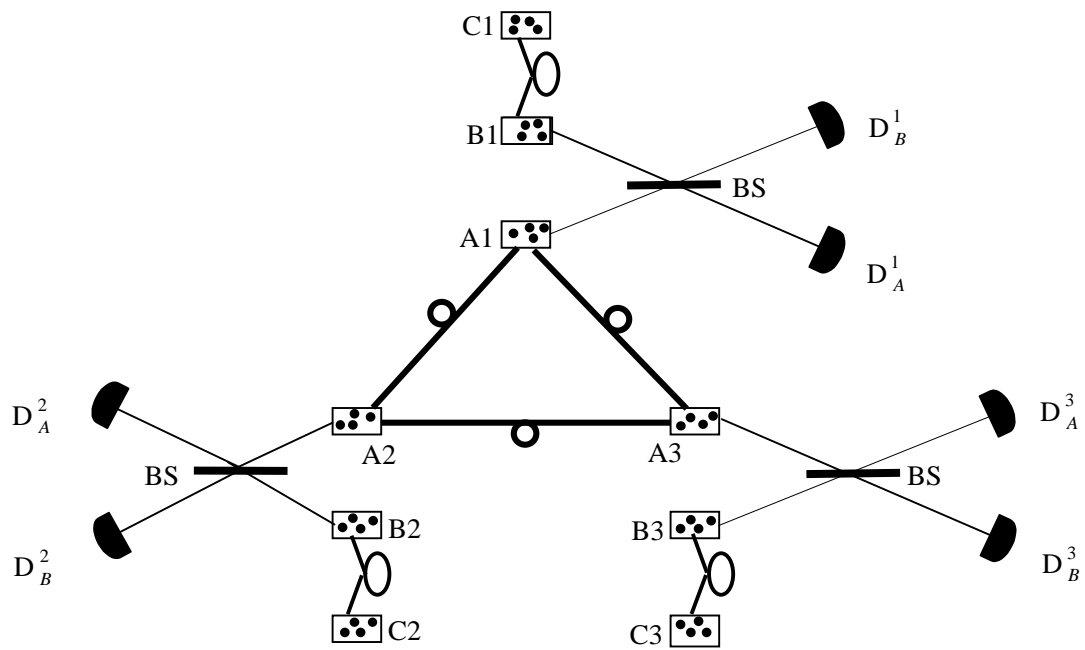
{\bf Figure 4. }Set-up for generating PME$\;W$ state between atomic
ensembles and demonstrating quantum nonlocality.\qquad

\end{document}